\documentclass[manuscript,screen,nonacm]{acmart}

\AtBeginDocument{%
  \providecommand\BibTeX{{%
    \normalfont B\kern-0.5em{\scshape i\kern-0.25em b}\kern-0.8em\TeX}}}

\begin{document}

\title{The Ethical Implications of AI in Creative Industries: A Focus on AI-Generated Art}


\author{Prerana Khatiwada}
\affiliation{%
  \institution{University of Delaware}
  % \streetaddress{1 Th{\o}rv{\"a}ld Circle}
  \city{Newark}
  \country{USA}
  }
\email{preranak@udel.edu
}
\author{Joshua Washington}
\affiliation{%
   \institution{University of Delaware}
  % \streetaddress{1 Th{\o}rv{\"a}ld Circle}
  \city{Newark}
  \country{USA}
  }
\email{joshwash@udel.edu}

\author{Tyler Walsh}
\affiliation{%
 \institution{University of Delaware}
  % \streetaddress{1 Th{\o}rv{\"a}ld Circle}
  \city{Newark}
  \country{USA}
  }
\email{tjwalsh@udel.edu}

\author{Ahmed Saif Hamed}
\affiliation{%
 \institution{University of Delaware}
  % \streetaddress{1 Th{\o}rv{\"a}ld Circle}
  \city{Newark}
  \country{USA}
  }
\email{ahmd@udel.edu}

\author{Lokesh Bhatta}
\affiliation{%
 \institution{Wilmington University}
  % \streetaddress{1 Th{\o}rv{\"a}ld Circle}
  \city{New Castle}
  \country{USA}
  }
\email{lbhatta001@my.wilmu.edu}

\renewcommand{\shortauthors}{Khatiwada et al.}

\begin{abstract}
As Artificial Intelligence (AI) continues to grow daily, more exciting (and somewhat controversial) technology emerges every other day. As we see the advancements in AI, we see more and more people becoming skeptical of it. This paper explores the complications and confusion around the ethics of generative AI art. 
We delve deep into the ethical side of AI, specifically generative art. We step back from the excitement and observe the impossible conundrums that this impressive technology produces. Covering environmental consequences, celebrity representation, intellectual property, deep fakes, and artist displacement. Our research found that generative AI art is responsible for increased carbon emissions, spreading misinformation, copyright infringement, unlawful depiction, and job displacement. In light of this, we propose multiple possible solutions for these problems. We address each situation's history, cause, and consequences and offer different viewpoints. At the root of it all, though, the central theme is that generative AI Art needs to be correctly legislated and regulated. 
\end{abstract}




\maketitle

\section{Introduction}
Many artists are not so thrilled at the idea that a machine might replace them one day \cite{verdegem2021ai, coeckelbergh2017can}.
This anxiety is obvious in the animation and digital art communities, where AI-generated content is rapidly gaining traction \cite{caporusso2023generative, lovato2024foregrounding}. 
A notable example is the high amount of backlash faced by the tech channel CorridorDigital \cite{jade}, where animation was made using machine learning software that greatly simplified the process \cite{5test}. 
While some praised the technological advancement, many artists criticized it as threatening their craft, creativity, and job security. Many people in the art field or planning to go into it are very concerned about the future, as it holds anything. 

Concern for the future doesn’t end here, as some artists are also in an uproar over the procurement of art for training data without consent—especially in cases like Corridor Digital\footnote{\url{https://www.pugetsystems.com/featured/case-study-with-corridor-digital/?srsltid=AfmBOoq_MCR-RaOTh1MoWcillHT_gQojU_GDNNjjysGJY5-XVV9xvV6l}} and ChatGPT.
This lack of transparency and respect for creative ownership raises serious questions about consent, attribution, and artistic integrity in the age of AI.

These concerns are not isolated incidents—they reflect a broader unease in the creative industries about the unchecked rise of generative AI \cite{amankwah2024impending}. One of the most pressing issues involves the unauthorized use of artists’ work to train large-scale machine learning models \cite{verdegem2021ai}. 
Many generative systems, such as Stable Diffusion \footnote{\url{https://stablediffusionweb.com/}} or Midjourney \footnote{\url{https://en.wikipedia.org/wiki/Midjourney}}, scrape images from the internet—including professional portfolios, illustrations, and concept art—without the consent or even awareness of the original creators.
This practice has sparked a wave of frustration and activism among artists, many of whom argue that their labor, style, and creative identity are being commodified without recognition or compensation \cite{kompatsiaris2015art}. But copyright infringement is just the beginning. As generative AI systems become more accessible and sophisticated, ethical dilemmas continue to multiply: Who owns AI-generated art? \cite{chesterman2023ai} Can a machine imitate the emotional intent of a human creator? \cite{demmer2023does} What happens when deepfakes and AI-generated images are used for deception or defamation in art, journalism, or activism? These aren’t the only concerns that have arisen, and probably won’t be the last ethical dilemmas. 
\looseness -1

The rise of AI-generated art has introduced a wave of ethical dilemmas that extend far beyond copyright questions. From unauthorized data scraping to stylistic mimicry, the creative industry is grappling with preserving artistic integrity in the face of rapid technological change. As these practices gain public visibility and legal scrutiny, calls for regulation are growing louder. Yet crafting legislation around AI art remains challenging, given the novelty and complexity of the issues at hand. The legal landscape remains fragmented and inconsistent across regions with generative tools evolving faster than policy can adapt. 

Some early efforts—such as the European Union’s AI Act \cite{europarl} and Japan’s debates around AI and copyright \cite{ueno2021flexible}—offer promising starting points, but globally, regulatory frameworks remain in their infancy. In this uncertain environment, ethical reasoning becomes a critical bridge, helping to guide decisions where law has yet to catch up. Ethical reasoning can help fill the gap between outdated laws and emerging realities, offering guidance on consent, attribution, fairness, and creative rights \cite{coke2017adapting}.
Although there is some iffiness about making laws around AI art since this is a relatively new issue that has sparked recently, and altogether, the emergence of AI art is very recent. So, if anything, this rise of ethical concerns calls for more ethical regulation and understanding \cite{coke2017adapting}, as ethics can help bridge the gap between the law and human morals. 

This paper explores the ethical implications of AI-generated art and examines the tensions between innovation and artistic rights. We aim to contribute to the growing discourse on how society might develop fair, transparent, and inclusive guidelines that ensure AI remains a tool for creativity, not a threat to it.

\section{Related Work}
This section reviews work on how machine learning is reshaping artistic creation and the ongoing efforts to establish ethical guidelines for AI-generated art. We examine the creative implications of AI in visual and narrative media, as well as emerging discussions around fairness, ownership, and artist rights in response to this evolving technology.
\looseness -1

\subsection{The Impact of Machine Learning on Art Creation} 
Over the past decade, numerous advancements have been made in AI, specifically machine learning \cite{coke2017adapting}. Those advancements have ranged drastically from computers being able to communicate with us to the automation of cars \cite{1test}. One field that has been significantly affected by machine learning is art \cite{4shoaib2023deepfakes}. This expressive field has been innovated in machine learning as programmers have implemented how machines can learn to make art. Machines learn how to make art by studying multiple images of art and then replicating aspects of each art piece to create \cite{4shoaib2023deepfakes}. For instance, a machine learning program to create an image of a dragon uses multiple images of dragons to create a unique image of a dragon. This isn’t limited to simple things either, as the machine learning software of today, like ChatGPT, can create images in any specific style, shape, or form based on any prompt a user gives \cite{1test}. These programs can create dozens of different images based on any font due to being fueled by thousands of other art images. With this advancement, one may think that people were ecstatic about this evolution in the art world. Unfortunately, many controversies have emerged from this newest innovation, including originality and authorship, devaluation of human creativity, ethical use of data, and erosion of artistic skills.

\subsection{Creating Fair Guidelines for A.I. Art and Innovation }
We have looked at both groups respectfully and have settled on rules we believe best suit users creating AI art or future artists, as well as the use of AI in general. In general, when making this rule set, we generally settled on the basis for which we would make rules that would allow further innovation to be made, as it is important for the future, in our opinion, if AI, even generative art, continues to grow. However, we also considered regulation in making sure people receive their credit if their work was being used, and if the artwork being generated was harmful in some way.

\section{Proposed Work--Discourse of Ethical Guidelines}
This section outlines our proposed approach to developing ethical guidelines for AI-generated art, focusing on transparency, artist rights, and accountability within creative industries.

\subsection{B1: Ethical Dilemma Around Carbon Emissions }
The first ethical dilemma we thought was the most important to discuss was the one around carbon emissions. Recently, technological advancements and regulations in the face of the pandemic have significantly minimized the amount of greenhouse gases produced \cite{8bolanvca2010carbon, 8test}. The reduction resulted from the need for more transportation as everyone went online \cite{8bolanvca2010carbon, 8test}. With this in mind, the reduction of greenhouse gases can be assisted by the introduction of artists using generative AI models for art. This addition will make creating the Artwork more convenient than it already is. The rise of digital art has paved the way and made creating art more and more convenient \cite{8bolanvca2010carbon, 8test}. The amount of pollution from supplies needed for the art and the reduction of fossil fuels used for transportation to get supplies \cite{8bolanvca2010carbon, 8test}.
Now that reducing carbon emissions from transportation is an issue that the convenience of AI can solve, there is still the issue of carbon emissions from the AI itself. Currently, the tracking and regulation of AI's carbon footprint, especially of massive learning models like ChatGPT, are nonexistent \cite{7brevini2021ai, 7test}. This is a pressing issue as they are proponents of carbon emissions. The issue becomes even more dire as temperatures climb, and recently, in 2023, it was marked as "the hottest year on record" \cite{7brevini2021ai, 7test}, meaning that regulation around the resource intensity of AI is needed more than ever. So, in response to this, we, as a society, need some ethical code or codes to minimize or completely erase the issue because, as of now, the AI models remain unethical as making these artworks lacks reliability and safety \cite{8bolanvca2010carbon, 8test}. So, AI should have a limit on how it is used by a person. This should solve the issue of carbon emissions as the people who used generative AI had a significantly lower impact on the environment when in a country that relies heavily on renewable energy, compared to countries that rely heavily on fossil fuels. The limit should then be heavier on individuals who live in countries with a majority of fossil fuels and, in contrast, be more lenient on individuals who live in countries that use a majority of renewable energy. This proposal will significantly reduce the carbon emissions from transportation in procuring art supplies, art supply waste in landfills, and overall waste from fossil-dependent countries. While we believe we solved this dilemma, a more technical one has arisen recently from generative AI art.

\subsection{B2: Ethical Dilemma Around Influence }
When scouring for ethical dilemmas that generative art faces, we encountered scandals related to the realism of the images being created. Continuing, we found a dilemma around falsified images being generated. And so, we decided that the next point we would tackle was AI's potential maliciousness. Currently, the amount of time to generate an image is minimal, and the amount of detail you can have is great. Even though this is a great innovation, some well-needed regulations regarding these machines depicting people are still needed. Currently, these machine learning models can create highly detailed pictures of celebrities, and with some input from the user, they could be made to make highly detailed pictures of people \cite{4test}. There is already devastating damage from false images of well-known people. Pictures have been of President Trump getting arrested, Biden cheating, or the Pope in a Balenciaga puffer jacket \cite{4test}. There is even a devastating photo of the Pentagon getting bombed, which caused over \$500 billion of US markets to be lost \cite{4test}. That is only part of it, however, as some of these images include videos using AI to imitate someone's voice. This technology is impressive, as in some cases, it is almost impossible to tell if the audio is real or generated by AI. While primarily used for comedic purposes, it begs the question of whether this type of technology is morally correct to produce and use. Thus, we decided that the most ethical way to handle AI-making images with personal people is either to have whatever image the AI creates, have some indication that it is false, have a blocklist of certain people when it comes to creating AI-generated art of them, or have a mixture of the two. Future researchers can research ways to automate software to search for and find deep fakes. Suppose the software comes upon a situation with this rule. In that case, it can be easier to identify generated images of art, so there are fewer fake images capable of causing lots of real-world damage, like the Pentagon bombing pictures. Then, the blocklist idea can be implemented so that it is harder for the fake images to surface and cause controversy. If Trump, the Pope, or Biden were blocklisted names to be used as prompts, then the deep fakes of Trump getting arrested, Biden cheating, or the Pope wearing a Balenciaga jacket would not make headlines as they did. Overall, these two rules, combined, will be the most ethical when it comes to generating images of others, and it is not the only rule we have left, either.

\subsection {B3: Ethical Dilemma Around Intellectual Property }
Intellectual property refers to patents, copyrights, and trademarks protected by law to give people recognition and financial rights to something they have created. This is a way to protect people's creations and allow them to claim them as their own \cite{12test}. This is especially important for those who wish to claim Artwork they have created as their own. It can not be distributed or recreated without the original artist's consent, so who gets credit whenever somebody generates a piece of art using AI software? Generative AI art software is still a new and impressive technology, allowing users to create detailed pieces of Artwork with just a text prompt. These images are made by using other art pieces, scraping them off the internet, and manipulating multiple art pieces to create a new one \cite{1test}. In the MIT Technology Review, Melissa Heikkilä writes about a Polish artist named Greg Rutkowski, who is well-known for his beautiful fantasy art. Because of his distinctive and detailed art style, he is subject to being used in many of the prompts, trying to imitate the kind of art he creates. Rutkowski is understandably upset, as his art is being used and practically stolen without his consent \cite{6rodrigues2020legal}—the rise of generative AI art affects those whose art is being used as training data for AI, but what about those who create the AI systems? Stephen Thaler created a generative AI art software under the alias DABUS.
After generating art, Thaler wanted the intellectual rights to the art, as the software he created created this piece of art. However, a federal judge declined his request, stating that it did not have the human authorship to be deemed his intellectual property \cite{9test}. This seems sorted. However, nobody gets any intellectual property, which could be an issue down the road. Even then, in the case with Thaler, there is still an ongoing lawsuit and discussion on AI art within the Copyright Office. Although he was denied the rights to the Artwork, it is not unlikely that he will eventually be granted ownership. As an attempt to mediate this issue, we offer some possible solutions. In this first outcome, nobody earns intellectual property from AI art. It is completely copyright-free and can be used freely. This has some flaws, mainly allowing anybody to profit from any AI-generated work, which is not right. Secondly, the person who created the original Artwork being used as training data sees no benefit and has to watch as their Artwork receives no recognition. This is the main issue, as the brilliant minds and talents behind the Artwork don't get any credit for their Artwork. This is why we see only two solutions as possible: nobody gets any credibility, and the AI Artwork can not be sold or used other than something that exists only to be created or for which the original author is credited. There should also be some legislation that monitors the type of training data used in these AI generative machines, which are used only with the original author's consent. That being said, this raises questions about the future employment and livelihood of current authors, as it suggests that in the near future, we will see a lack of need for artists to create anything other than training data for AI, which brings us to our next point. 

\subsection{B4: Ethical Dilemma Around Deep Fakes }
One of the most significant problems with AI is deepfakes. As the word indicates, a deepfake is artificially generated content, primarily videos, by a machine learning model often deemed as "deep" \cite{14westerlund2019emergence}. The amount of deepfake content on the internet has reached alarming numbers. With the widespread accessibility of the internet, tasks once considered impossible or difficult are now achievable by almost any user. For example, video editing has become straightforward with just a few clicks, thanks to GPT websites, leading to the concerning problem of deepfakes. As of 2023, the accessible use of AI tools has increased deepfake content, rising more than 550\% from 2019, as reported by the Home Security Heroes team. Not only that, 98\% of online deepfake videos are pornography. This discussion brings us back to the case of Rana Ayyub, who was a victim of deepfake pornography, significantly impacting her mental well-being and affecting her family. The harm ruined her mental state and even reached her family. Another example of deepfakes is a viral video of Hollywood actor Leonardo DiCaprio giving a speech at the United Nations Climate Summit \cite{17testttt}. What is strange about the clip is that DiCaprio speaks with different famous voices, such as Joe Rogan, Steve Jobs, Robert Downey Jr, and many more, with perfect lip movement. With all that in mind, this unprecedented super-powerful technology thus challenges the concept of truth in digital media, complicating the distinction between authentic and manipulated content. If this issue is not tackled appropriately, we might see national leaders declaring war on one another in a random TikTok video. Therefore, a set of guidelines and steps must be provided and followed by both the government and the citizens to minimize the harm of deepfakes. Technologically, developing and deploying detection tools that use AI to differentiate between genuine and manipulated content can help identify deepfakes. Instead of IT scientists setting AI tools, we can have researchers do the opposite and create deepfake detection tools. For the legal side, implementing stricter regulations and laws to penalize the creation and distribution of harmful deepfake content can deter malicious use. 
\looseness -1

\subsection{B5: The Ethical Dilemma Around Displacement}
The revolution of AI-generated art presents a significant shift in the creative landscape, intertwining with the fabric of traditional artistry and introducing complex ethical and economic considerations. Integrating AI into the art world does not merely present a new medium for expression, but also raises essential questions about the future job market for traditional artists. As AI-generated artworks become more dominant, the art community faces the dual challenge of embracing technological innovation while safeguarding the livelihoods of human artists. Moreover, integrating AI into art creation without carefully crafted guidelines risks undermining human artists' economic stability and cultural significance. Such guidelines could serve as a safeguard, ensuring that the burgeoning realm of AI art does not diminish the value of human creativity but rather complements it, fostering a collaborative rather than competitive relationship. However, in his article, Krzysztof Pelc proposes an opposite view \cite{13horton2023bias}. He suggests that as AI-generated art becomes more common, societal tastes will evolve to prize more human-created art, eliminating all concerns about human-made art's future. Pelc also points out that the unique qualities of human art, tied to personal vision and passion, will become increasingly valued, making human art more precious in contrast to AI-generated works. 

Accessibility and inclusivity emerge as central ethical concerns around experts in AI-generated art. Ensuring this new art form does not perpetuate existing inequalities or exclude underrepresented groups requires a concerted effort to democratize access to AI technologies and promote diversity within the datasets that train these models. Collaborative efforts between human artists and AI further highlight the potential for a harmonious blending of creativity and technology, challenging us to redefine the boundaries of art. Such collaborations should be navigated with an ethical compass that respects the contributions of both parties, fostering an environment where human creativity and AI augmentation enhance one another in the pursuit of artistic innovation. Following this narrative, the relationship between human creativity and AI augmentation must be carefully calibrated to ensure it remains ethically balanced, honoring the intrinsic value of human artistic expression while exploring the boundless possibilities offered by AI \cite{11testttt}. By addressing these multifaceted concerns through thoughtful regulation, respectful cultural engagement, and inclusive policies, the art world can navigate the complexities introduced by AI-generated art, ensuring that this technological evolution enriches the artistic landscape rather than diminishes it. 

\section{CONCLUSION AND FUTURE WORK}
Overall, these codes we introduced should be sufficient in regulating carbon emissions of generating AI, settling copyright concerns, calming concerns of replacing artists, and controlling deep fakes. With our ethical codes for carbon emissions, it should prove safer to have everyone limited to the amount of carbon emissions they produce. The emergence of deep fake detection tools can help mitigate the spread of false information about deepfakes. Also, giving credit and potential compensation to any author of an artwork that was nonconsensually used will help maintain an ethical system of producing art fairly, conveniently, and inclusively. And overall, with artists continuing to generate art from artificial intelligence, the value of human art will remain the same, if not increased. Not just people in the art world but society as a whole can be greatly contributed to by AI art, and these new regulations and codes would prove to be a more ethical means of creation. 
One significant limitation of our study is the need for real-world testing for the four dilemmas we addressed. The hypotheses we propose remain largely hypothetical, needing more empirical data to solidify the effectiveness of our ethical codes. Moving forward, it is essential to explore how AI can contribute to reducing carbon emissions, especially in a global shift towards renewable energy sources. This exploration could improve the benefits of generative art. Furthermore, we did not examine the energy consumption of digital art versus the creation of art by a large generative model. This aspect deserves focused attention for a more comprehensive understanding and robust research outcomes. 

\section{Acknowledgments}
This work was conducted as part of the Computer Ethics and Society course at the University of Delaware. We want to thank Dr. Matthew Louis Mauriello for his valuable early feedback and thoughtful guidance during the development of this paper. We also thank our classmates for their insightful comments and suggestions during in-class discussions, which helped shape and refine our ideas.

\bibliographystyle{ACM-Reference-Format}
\bibliography{software}

\end{document}